\documentclass[a4paper,11pt]{article}
\usepackage{pos}
\usepackage{subcaption}
\usepackage{amsmath}
\usepackage{bigints}
\usepackage{mathrsfs}
\usepackage{pgfplots}
\usepackage{multicol}
\usepackage{verbatim}
\usepackage[T1]{fontenc}
\usepackage{slashed}
\usepackage[utf8]{inputenc}
\usepackage{array}
\usepackage{wrapfig}
\usepackage{multirow}
\usepackage{tabu}
\usepackage{tabularx}
\usepackage{tabularray}
\usepackage{amsmath}
\usepackage{amsthm}
\usepackage{enumitem}
\usepackage{graphicx}
\usepackage{caption}
\usepackage{placeins}
\usepackage{longtable}
\usepackage{float}
\usepackage{amsfonts}
\usepackage{caption}
\usepackage{subcaption}
\usepackage{tikz,tikz-3dplot}
\usepackage[compat=1.0.0]{tikz-feynman}
\usepackage{axodraw2}
\usetikzlibrary{calc,arrows.meta,shapes.geometric}
\usepackage{soul}
\newcommand{\pysecdec}{py{\textsc{SecDec}}}

\title{Evaluating Parametric Integrals in the Minkowski Regime without Contour Deformation}
\author[a]{Stephen Jones}
\author[b]{Anton Olsson}
\author*[a]{Thomas Stone}

\affiliation[a]{Institute for Particle Physics Phenomenology (IPPP), Durham University,\\Durham, UK}
\affiliation[b]{Institute for Theoretical Physics, Karlsruhe Institute of Technology (KIT),\\76131 Karlsruhe, Germany}

\emailAdd{stephen.jones@durham.ac.uk}
\emailAdd{anton.olsson@kit.edu}
\emailAdd{thomas.w.stone@durham.ac.uk}

\abstract{We present selected examples demonstrating an alternative approach to contour deformation for numerically computing loop integrals in the Minkowski regime. This method focuses on identifying singular hypersurfaces (varieties of the $\mathscr{F}$ polynomial) and mapping them to known points which can then be resolved by employing blow-ups/sector decomposition techniques, thereby avoiding the need for contour deformation. Using this technique, we achieve improved convergence properties without the need for contour deformation, which is known to significantly increase the complexity of the integrand by introducing, for example, derivatives of the $\mathscr{F}$ polynomial and complicated Jacobians. We highlight that while we have only tested the approach on selected one-, two- and three-loop massless and one-loop massive examples, it shows promise for practical applications, offering potential benefits over the traditional approach. Evaluation times are compared with existing contour deformation implementations to illustrate the performance of this alternative method.}

\FullConference{Loops and Legs in Quantum Field Theory (LL2024)\\
 14-19, April, 2024\\
Wittenberg, Germany\\}

\begin{document}
\maketitle

\section{Introduction \& Motivation}
Many loop integrals appearing in state-of-the-art amplitude calculations are analytically intractable; as a result of this, numerical methods and other approximations have been developed to tackle these integrals \cite{Heinrich_2024,Smirnov_2022,Borinsky_2023,capatti2020manifestlycausallooptreeduality,bobadilla2021lottylooptreeduality,Czakon_2006,Gluza:2007rt,Liu_2023,hidding2020diffexpmathematicapackagecomputing,Armadillo_2023,Borowka_2018}. Even with these tools, we often have trouble numerically calculating in the
so-called ``Minkowski'' regime, which we define below, due to poles on the contour of integration. To remedy this, we deform the contour of integration away from the real axis to avoid these poles in such a way that we respect the causal $i\delta$ (Feynman) prescription. Methods have been explored to remove the need for contour deformation for momentum-space integrals (for example, in the context of loop-tree duality \cite{buchta2014looptreedualitywork,Capatti_2020,Kermanschah_2022}) so it is a natural question to ask whether it is possible to do the same in Feynman parameter space. 

As a reminder, we show how an integral in momentum space appears once it has been cast into Feynman-parameterised form:
\begin{equation}
\begin{gathered}
I=\bigintsss\limits_{-\infty}^{+\infty}\left(\prod\limits_{l=1}^{L} \frac{\mathrm{d}^{D}k_{l}}{i \pi^{\frac{D}{2}}}\right)\prod\limits_{j=1}^{N} \frac{1}{P^{\nu_{j}}_{j}\left(\{k\},\{p\},m_{j}^{2}\right)}\\\Downarrow\\
I=\frac{\left(-1\right)^{\nu}\Gamma\left(\nu-L D/2\right)}{\prod_{j=1}^{N}\Gamma\left(\nu_{j}\right)}\!\bigintsss\limits_{\mathbb{R}_{\geq0}^{N}}\left(\prod\limits_{j=1}^{N}\mathrm{d}x_{j} x_{j}^{\nu_{j}-1}\right)\frac{\mathscr{U}\!\left(\mathbf{x}\right)^{\nu-(L+1)D/2}}{\left(\mathscr{F}\!\left(\mathbf{x},\mathbf{s}\right)-i\delta\right)^{\nu-LD/2}}\delta\left(1-\sum\limits_{j=1}^{N}x_j\right)
\end{gathered}
\end{equation}
where $D$ is the spacetime dimension, $L$ is the number of loops and $N$ is the number of propagators (denoted by $P_j$ which are of the standard ``$p^2-m^2+i\delta$'' momentum-space form) and we allow for dots ($\nu_j\in\mathbb{N}$). After Feynman parameterisation, we have an integral of the Feynman parameters $\{x_j\}$ over the positive hyper-quadrant $\mathbb{R}_{\geq0}^{N}$ with an integrand containing the homogeneous Symanzik polynomials $\mathscr{U}\!\left(\mathbf{x}\right)$ and $\mathscr{F}\!\left(\mathbf{x},\mathbf{s}\right)$ of degree $L$ and $L+1$ respectively.

We note that there are several equivalent algorithms to construct the Symanzik polynomials, including constructing directly from cutting propagators in the Feynman diagram associated to the relevant integral (where this diagram exists). It is also instructive to emphasise that $\mathscr{F}$ is dependent on the kinematics of the problem (denoted by $\mathbf{s}$) whereas $\mathscr{U}$ depends solely on the Feynman parameters and is strictly non-negative in the integration domain (and hence, does not need an $i\delta$ prescription). We define the (pseudo-)Euclidean regime to be the region of kinematic space where $\mathscr{F}\!\left(\mathbf{x},\mathbf{s}\right)\geq0$ everywhere in the domain of integration (vanishing only on the boundary) and the Minkowski regime is henceforth defined to be the remaining region of kinematic space which is \textbf{not} in the (pseudo-)Euclidean regime (e.g. $\mathcal{M}=\{(s,t)\in\mathbb{R}^2\ |\ s>|t|\land t<0\}$ for on-shell massless $2\to2$ scattering). In this region, $\mathscr{F}$ can take different signs within the domain of integration and hence, in this regime, a prescription is required to deform the contour in a way that obeys causality. The causal $i\delta$ prescription in momentum space induces $\mathscr{F}-i\delta$ in Feynman parameter space and this is needed when $\mathscr{F}=0$ \textit{within} the domain of integration (that is to say, \textit{within} $\mathbb{R}_{>0}^{N}$ but strictly \textit{not} on the boundary).

The standard way this is implemented in Feynman parameter space is as follows \cite{PhysRevD.62.014009,Binoth_2005,PhysRevLett.100.241806,BOROWKA2013396}:
\begin{equation}\label{deffeynpar}
\mathscr{F}\left(\mathbf{z}\right)=\mathscr{F}\left(\mathbf{x}\right)-i\sum\limits_{j}\tau_{j}\frac{\partial\mathscr{F}\left(\mathbf{x}\right)}{\partial x_{j}},\qquad\tau_j=\lambda_j x_j \left(1-x_j\right) \frac{\partial\mathscr{F}\left(\mathbf{x}\right)}{\partial x_{j}}
\end{equation}
such that $\mathscr{F}$ is given a small imaginary part with the deformation away from the real axis parameterised by $\{\lambda_j\}$. Many techniques, including neural networks \cite{Winterhalder_2022}, have been used to optimise this choice of $\{\lambda_j\}$ but overall this can still slow down numerical integration massively (due to increased variance of the integrand leading to cancellations between postive and negative quantities of similarly large magnitude). There are even cases where this procedure fails entirely and this is related to the presence of a leading Landau singularity for arbitrary kinematics (see, for example, \cite{sjpaper}). This happens when, for all kinematic points in the Minkowski regime, $\mathcal{M}$, there exists (at least) one point where $\mathscr{F}\left(\mathbf{x}\right)$ and all $\frac{\partial\mathscr{F}\left(\mathbf{x}\right)}{\partial x_{j}}$ are simultaneously zero so the construction in Eq.~\ref{deffeynpar} breaks down. We encounter an example of this in section~\ref{sec:3loopsec}.

For the reasons described above, it would be beneficial to avoid this procedure of contour deformation in Feynman parameter space where possible; that motivation is the focus of the work presented in these proceedings.

\subsection{The Idea}
\begin{figure}[t]
    \centering
    \raisebox{-0.5\height}{\includegraphics[width=0.45\textwidth]{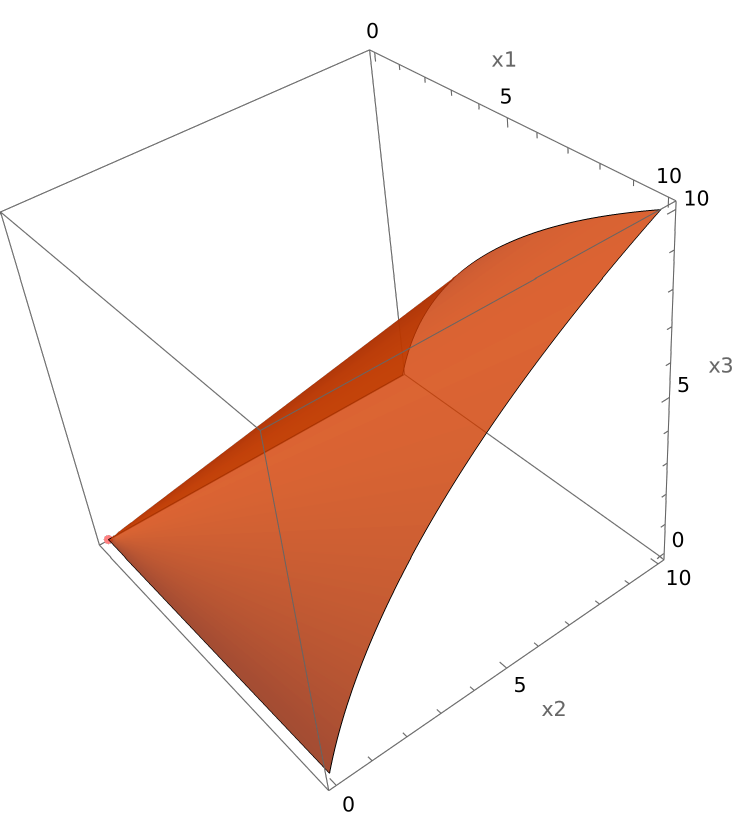}}
    \hfill $\Rightarrow$ \hfill
    \raisebox{-0.5\height}{\includegraphics[width=0.45\textwidth]{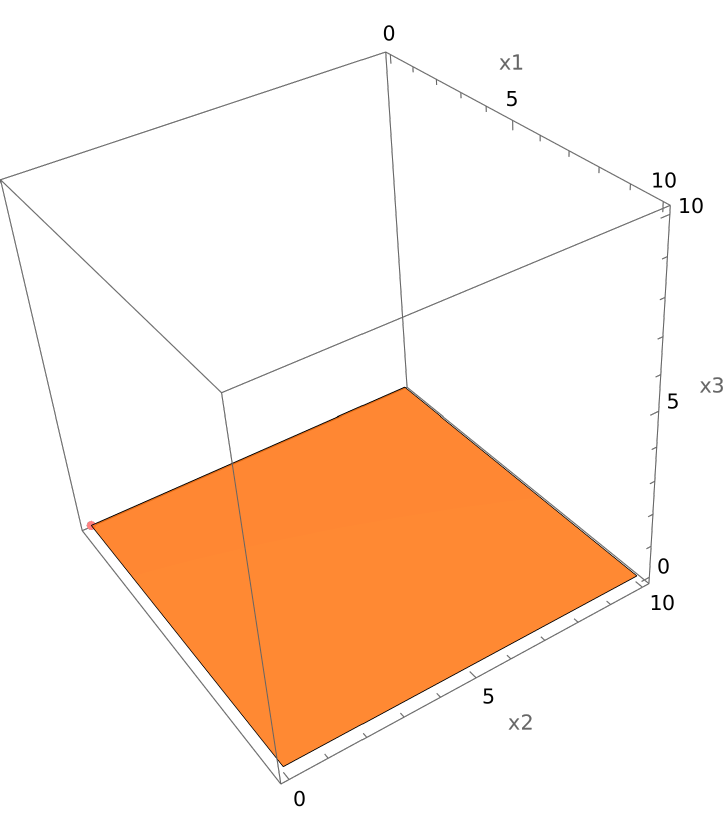}} \\
    \caption{A one-loop example hypersurface defined by $\mathscr{F}=0$ before (\textbf{left}) and after (\textbf{right}) transformation.}
    \label{fig:fpolytransform}
\end{figure}

The central idea is to construct transformations of the Feynman parameters $\{x_j\}$ which map zeroes of the $\mathscr{F}$ polynomial to the boundary of integration as illustrated in an example in Fig.~\ref{fig:fpolytransform}.
%
%
%\noindent\\
%\begin{minipage}{0.4\textwidth}
%    \includegraphics[width=\textwidth]{curvetriangle1.pdf}
%\end{minipage}
%\hfill
%\begin{minipage}{0.1\textwidth}
%    \begin{center}
%        $\Rightarrow$
%    \end{center}
%\end{minipage}
%\hfill
%\begin{minipage}{0.4\textwidth}
%    \includegraphics[width=\textwidth]{curvetriangle2.pdf}
%\end{minipage}
%\\\vspace{0.1cm}\\
\noindent These transformations look like a) positive parameter rescalings (e.g. $x_j\rightarrow \alpha x_j$ or $x_j\rightarrow x_i x_j$) and b) the introduction of hierarchies between Feynman parameters which generate split integrals to cover the entire original parameter space (e.g. $x_j\rightarrow x_i+x_j$). Transformations of type a) are justified by the projective invariance of the integral and the Cheng-Wu theorem \cite{Cheng:1987ga,panzerthesis,weinzierl2022feynmanintegrals} and transformations of type b) are justified by inserting the Heaviside identity, $\theta\left(x_{a}-x_{b}\right)+\theta\left(x_{b}-x_{a}\right)=1$, under the integral sign. These transformations are very reminiscent of sector decomposition \cite{HEINRICH_2008} (and indeed, this work is greatly inspired by that technique) but we stress that sector decomposition as used for the numerical computation of Feynman integrals in the physics literature is a method for dealing with singularities on the boundary of the domain of integration whereas this technique maps singularities \textit{within} the domain \textit{to} the boundary where they can then be dealt with using the sector decomposition algorithm (for which we use \pysecdec{} \cite{Heinrich_2024}).

The result of these transformations is that we only ever integrate manifestly non-negative integrands (for the transformations which make $\mathscr{F}$ non-positive, we extract an overall minus sign from $\mathscr{F}$ and factor it outside the integral along with the $i\delta$ all to the appropriate power which generates the physically-correct imaginary part). We stitch everything together as follows:
\begin{equation}
I=\sum\limits_{n_+=1}^{N_+}I^{+}_{n_+}+(-1-i\delta)^{-\left(\nu-LD/2\right)}\sum\limits_{n_-=1}^{N_-}I^{-}_{n_-}
\end{equation}
where $I^{+(\!-\!)}_{n_{+(\!-\!)}}$ corresponds to an integral where the transformed $\mathscr{F}$ polynomial is non-negative (non-positive). As we will demonstrate in a number of specific examples, it can be \textbf{much} faster to numerically integrate the manifestly non-negative integrands of the integrals $\{I^{+}_{n_+},I^{-}_{n_-}\}$. We begin by considering a selection of integrals with massless internal lines.

\section{Massless Integrals}
\begin{figure}[t]
    \centering
    \begin{subfigure}[t]{0.33\textwidth}
        \centering
        \begin{tikzpicture}[baseline=13ex,scale=1.0]
%\draw [help lines] (0,0) grid (4,4);
%coordinates:
\coordinate (x1) at (1, 3) ;
\coordinate (x2) at (1, 1) ;
\coordinate (x3) at (3, 3) ;
\coordinate (x4) at (3, 1) ;
%external momenta:
\node (p1) at (0, 3) {$p_1$};
\node (p2) at (0, 1) {$p_2$};
\node (p3) at (4, 3) {$p_3$};
\node (p4) at (4, 1) {$p_4$};
%jets:
\draw[ultra thick,color=ForestGreen] (x1) -- (p1);
\draw[color=blue] (x2) -- (p2);
\draw[color=blue] (x3) -- (p3);
\draw[color=blue] (x4) -- (p4);
%edges:
\draw[ultra thick,color=Black] (x1) -- (x2) node [midway,xshift=-10pt,color=Red] {\Large $x_1$};
\draw[ultra thick,color=Black] (x1) -- (x3) node [midway,yshift=+10pt,color=Red] {\Large $x_0$};
\draw[ultra thick,color=Black] (x2) -- (x4) node [midway,yshift=-10pt,color=Red] {\Large $x_2$};
\draw[ultra thick,color=Black] (x3) -- (x4) node [midway,xshift=+10pt,color=Red] {\Large $x_3$};
%vertices:
\draw[fill,thick,color=Blue] (x1) circle (1pt);
\draw[fill,thick,color=Blue] (x2) circle (1pt);
\draw[fill,thick,color=Blue] (x3) circle (1pt);
\draw[fill,thick,color=Blue] (x4) circle (1pt);
        \end{tikzpicture}
    \end{subfigure}
    \hfill
    \begin{subfigure}[t]{0.33\textwidth}
        \centering
        \begin{tikzpicture}[baseline=13ex,scale=1.0]
%\draw [help lines] (0,0) grid (4,4);
%coordinates:
\coordinate (x1) at (1, 3) ;
\coordinate (x2) at (1, 1) ;
\coordinate (x3) at (2, 2) ;
\coordinate (x5) at (3, 1) ;
\coordinate (x4) at (3, 3) ;
%external momenta:
\node (p1) at (0, 3) {$p_1$};
\node (p2) at (0, 1) {$p_2$};
\node (p3) at (1.3, 1.3) {$p_3$};
\node (p4) at (4, 3) {$p_4$};
%jets:
\draw[color=blue] (x1) -- (p1);
\draw[color=blue] (x2) -- (p2);
\draw[color=blue] (x3) -- (p3);
\draw[color=blue] (x4) -- (p4);
%edges:
\draw[ultra thick,color=Black] (x1) -- (x2);
\draw[ultra thick,color=Black] (x1) -- (x4);
\draw[ultra thick,color=Black] (x2) -- (x5);
\draw[ultra thick,color=Black] (x3) -- (x1) ;
\draw[ultra thick,color=Black] (x3) -- (x5);
\draw[ultra thick,color=Black] (x4) -- (x5);
%vertices:
\draw[fill,thick,color=Blue] (x1) circle (1pt);
\draw[fill,thick,color=Blue] (x2) circle (1pt);
\draw[fill,thick,color=Blue] (x3) circle (1pt);
\draw[fill,thick,color=Blue] (x4) circle (1pt);
\draw[fill,thick,color=Blue] (x5) circle (1pt);
        \end{tikzpicture}
    \end{subfigure}
    \hfill
    \begin{subfigure}[t]{0.32\textwidth}
        \centering
        \begin{tikzpicture}[baseline=12ex,scale=0.35]

\draw (0.5,1) edge [color=blue] (2,2) node [] {};
\draw (9.5,1) edge [color=blue] (8,2) node [] {};
\draw (0.5,9) edge [color=blue] (2,8) node [] {};
\draw (9.5,9) edge [color=blue] (8,8) node [] {};
\draw (5,7.5) edge [ultra thick, bend right = 10] (2,2) node [] {};
\draw (5,7.5) edge [ultra thick, bend left = 10] (8,2) node [] {};
\draw (5,7.5) edge [ultra thick, bend right = 10] (2,8) node [] {};
\draw (5,7.5) edge [ultra thick, bend left = 10] (8,8) node [] {};
\draw (5,2.5) edge [ultra thick, draw=white, double=white, double distance=3pt, bend left = 10] (2,2) node [] {};\draw (5,2.5) edge [ultra thick, bend left = 10] (2,2) node [] {};
\draw (5,2.5) edge [ultra thick, draw=white, double=white, double distance=3pt, bend right = 10] (8,2) node [] {};\draw (5,2.5) edge [ultra thick, bend right = 10] (8,2) node [] {};
\draw (5,2.5) edge [ultra thick, draw=white, double=white, double distance=3pt, bend left = 10] (2,8) node [] {};\draw (5,2.5) edge [ultra thick, bend left = 10] (2,8) node [] {};
\draw (5,2.5) edge [ultra thick, draw=white, double=white, double distance=3pt, bend right = 10] (8,8) node [] {};\draw (5,2.5) edge [ultra thick, bend right = 10] (8,8) node [] {};

\node () at (0.3,0.5) {$p_1$};
\node () at (9.7,0.5) {$p_3$};
\node () at (0.3,9.5) {$p_2$};
\node () at (9.7,9.5) {$p_4$};

\draw[fill, thick, color=Blue] (2,2) circle (6pt);
\draw[fill, thick, color=Blue] (8,2) circle (6pt);
\draw[fill, thick, color=Blue] (2,8) circle (6pt);
\draw[fill, thick, color=Blue] (8,8) circle (6pt);
\draw[fill, thick, color=Blue] (5,7.5) circle (6pt);
\draw[fill, thick, color=Blue] (5,2.5) circle (6pt);
        \end{tikzpicture}
    \end{subfigure}
 \caption{A selection of massless integrals to which the method has been applied. Thin blue lines denote on-shell external legs, thick green lines denote off-shell external legs and black lines denote massless internal propagators. For the one-loop example, we also explicitly associate Feynman parameters (in red) to each propagator to simplify the connection between the diagram and its cut-constructed $\mathscr{U}$ and $\mathscr{F}$ polynomials.}
    \label{fig:masslessdiags}
\end{figure}
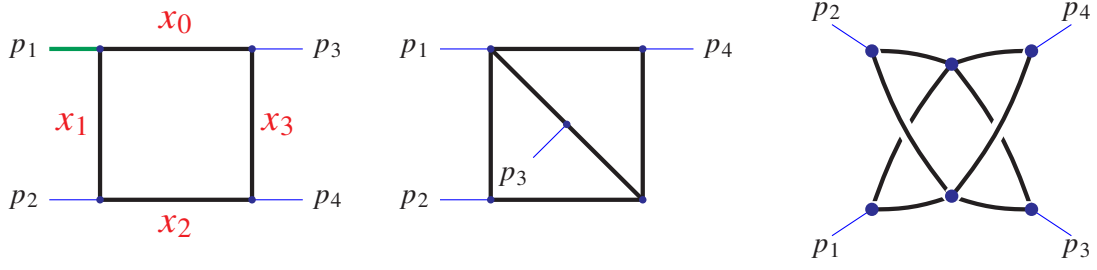

Our improvements on timings for numerical integration are best illustrated in our application of the method to massless integrals (by which we refer to integrals whose internal propagators have no masses; however, we do allow for off-shell external legs). To make the technique concrete, we will explicitly go through a simple one-loop massless box and then demonstrate the efficiency increases on two-loop and three-loop cases. The Feynman diagrams of the specific example integrals we consider are shown in Fig.~\ref{fig:masslessdiags}.

\subsection{One-Loop Off-Shell Box}
We begin by considering the one-loop off-shell box. For this example, 
\begin{align}
    \mathscr{U}&=x_{0}+x_{1}+x_{2}+x_{3}\\\mathscr{F}&=-s x_{0}x_{2}-t x_{1}x_{3}-p_{1}^{2}x_{0}x_{1}
\end{align} where $s$ and $t$ are the usual Mandelstam variables.

\begin{figure}[t]
    \centering
    \includegraphics[width=1.0\textwidth]{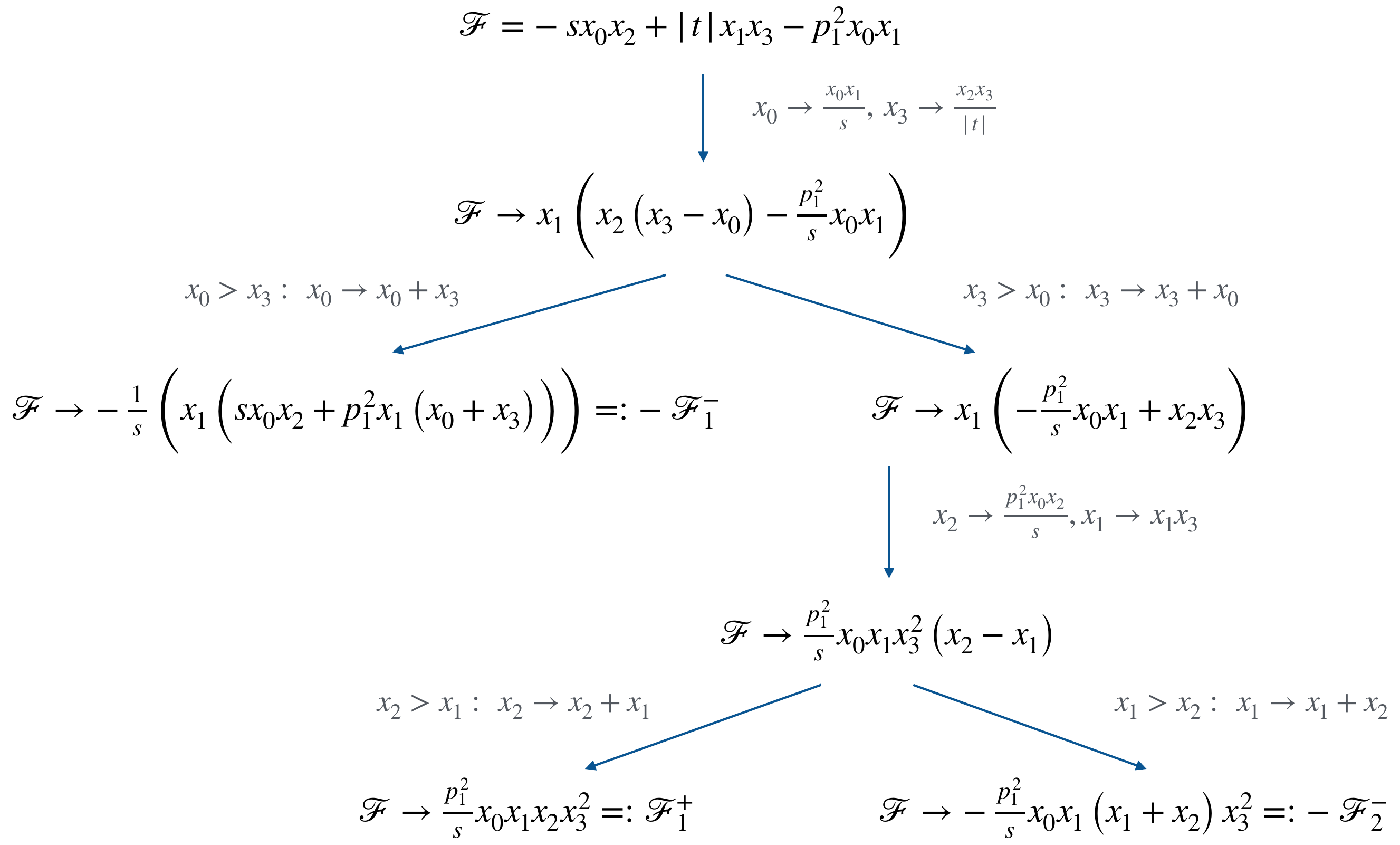}
    \caption{A flowchart depicting the series of transformations which resolves the singularities for the one-loop off-shell box.}
    \label{fig:flowchart}
\end{figure}

If we restrict to ${\mathcal{M}_{\mathrm{phys}}=\{(s,t,p_1^2)\in\mathbb{R}^3\ |\ s>0\land t<0\land p_1^2>0\}}$, there will be zeroes of $\mathscr{F}$ \textit{within} the integration volume for $(x_{0},x_{1},x_{2},x_{3})\in\mathbb{R}_{>0}^{4}$. We show in Fig.~\ref{fig:flowchart} the series of transformations which resolves the singularities due to $\mathscr{F}$.
\noindent To be consistent, we must also generate $\mathscr{U}^+_1$, $\mathscr{U}^-_1$ and $\mathscr{U}^-_2$ by applying the corresponding transformations to $\mathscr{U}$ as well as generating the absolute values of the corresponding Jacobian determinants: $\mathscr{J}^+_1$, $\mathscr{J}^-_1$ and $\mathscr{J}^-_2$. 

Having calculated these objects, we have a set of manifestly non-negative integrands with the same structure, $\ \sim\!\!\!\mathscr{J}^\pm_{n_\pm}\left(\mathscr{U}^\pm_{n_\pm}\right)^{2\epsilon}\!\left(\mathscr{F}^\pm_{n_\pm}\right)^{-2-\epsilon}$, which we can combine to express our initial integral into the following sum of three easier integrals:
\begin{equation}
    I=I_1^++(-1-i\delta)^{-2-\epsilon}\left(I^-_1+I^-_2\right)
\end{equation}
and we have verified this construction numerically integrates to the known analytic result using \pysecdec.

\subsection{Two-Loop Non-Planar Box}
\begin{figure}[t]
      \centering
      \begin{subfigure}{0.49\textwidth}
        \includegraphics[width=1.0\textwidth]{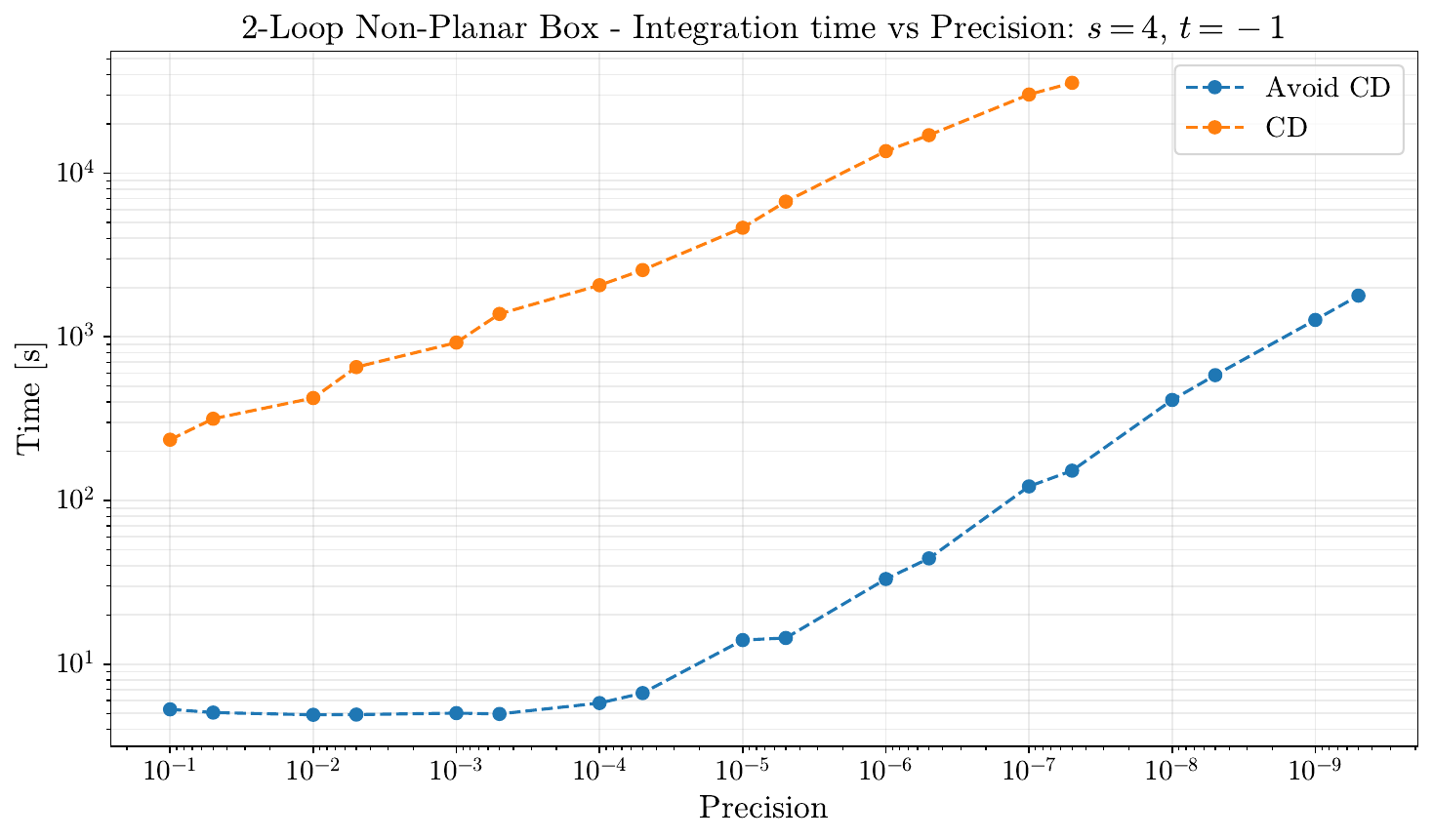}\caption{A kinematic point in $\mathcal{M}_1$}
\label{fig:subim1}  
      \end{subfigure}
      \begin{subfigure}{0.49\textwidth}
          \includegraphics[width=1.0\textwidth]{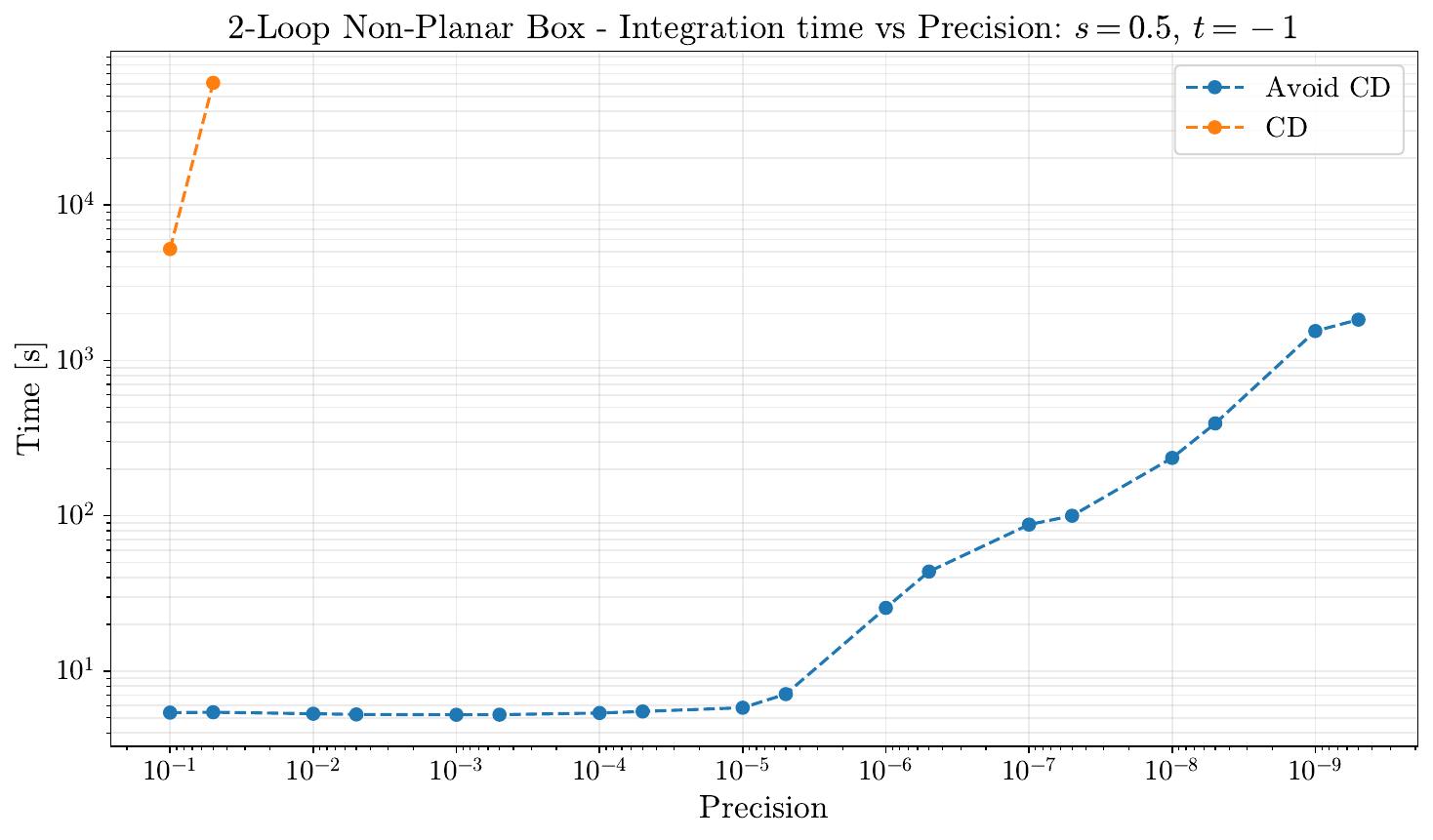}
          \caption{A kinematic point in $\mathcal{M}_2$}
\label{fig:subim2}
      \end{subfigure}

      \caption{Timing vs. precision comparisons avoiding contour deformation for the two-loop non-planar box. Orders of magnitude improvements are observed for points in both $\mathcal{M}_1$ (\textbf{a}) and $\mathcal{M}_2$ (\textbf{b}).}
      \label{fig:2loopplots}
  \end{figure}

We now consider a slightly more complicated example: a two-loop non-planar box. After cut-construction, we find that $\mathscr{F}=-s x_1 x_2 x_5 -t x_0 x_1 x_3-u x_0 x_2 x_4$ and furthermore, applying the momentum conservation constraint $s+t+u=0$ allows us to eliminate $u$ resulting in
\begin{equation}
    \mathscr{F}=-s x_1 x_2 x_5 -t x_0 x_1 x_3+(s+t) x_0 x_2 x_4.
\end{equation}
It is clear to see that even with $s$ and $t$ of the same sign, $\mathscr{F}$ can be zero \textit{within} $\mathbb{R}_{>0}^{6}$ and, in fact, a (pseudo-\nolinebreak)Euclidean region does not exist for this integral -- the entire kinematic space is in the Minkowski regime. To test our method, we consider two distinct subsets of the Minkwoski regime:
\begin{align}
    &\mathcal{M}_1:=\{\ (s,t)\in\mathbb{R}^2\ |\ s>|t|\ \land\ t<0\ \}\\
    &\mathcal{M}_2:=\{\ (s,t)\in\mathbb{R}^2\ |\ 0<s<|t|\ \land\ t<0\ \}.
\end{align}
We find that, to cover all the space in the regime defined by $\mathcal{M}_1$, we need six integrals combined in the form:
\begin{equation}
    I=\left(I_1^++I_2^++I_3^+\right)+(-1-i\delta)^{-2-2\epsilon}\left(I^-_1+I^-_2+I^-_3\right)
\end{equation}
which agrees with the analytic result \cite{Tausk_1999} when integrated numerically. For the regime $\mathcal{M}_2$, we similarly find a total of six (different) integrals. We compare the timings for integrating up-to-and-including finite order in $\epsilon$ with the standard contour deformation setup in \pysecdec{} versus our method in Fig.~\ref{fig:2loopplots} and we find significant improvements for a point in $\mathcal{M}_1$ (\ref{fig:subim1}) and $\mathcal{M}_2$ (\ref{fig:subim2}).\\

\subsection{Three-Loop Non-Planar Box}\label{sec:3loopsec}
\begin{figure}[t]
    \centering
    \includegraphics[width=0.9\textwidth]{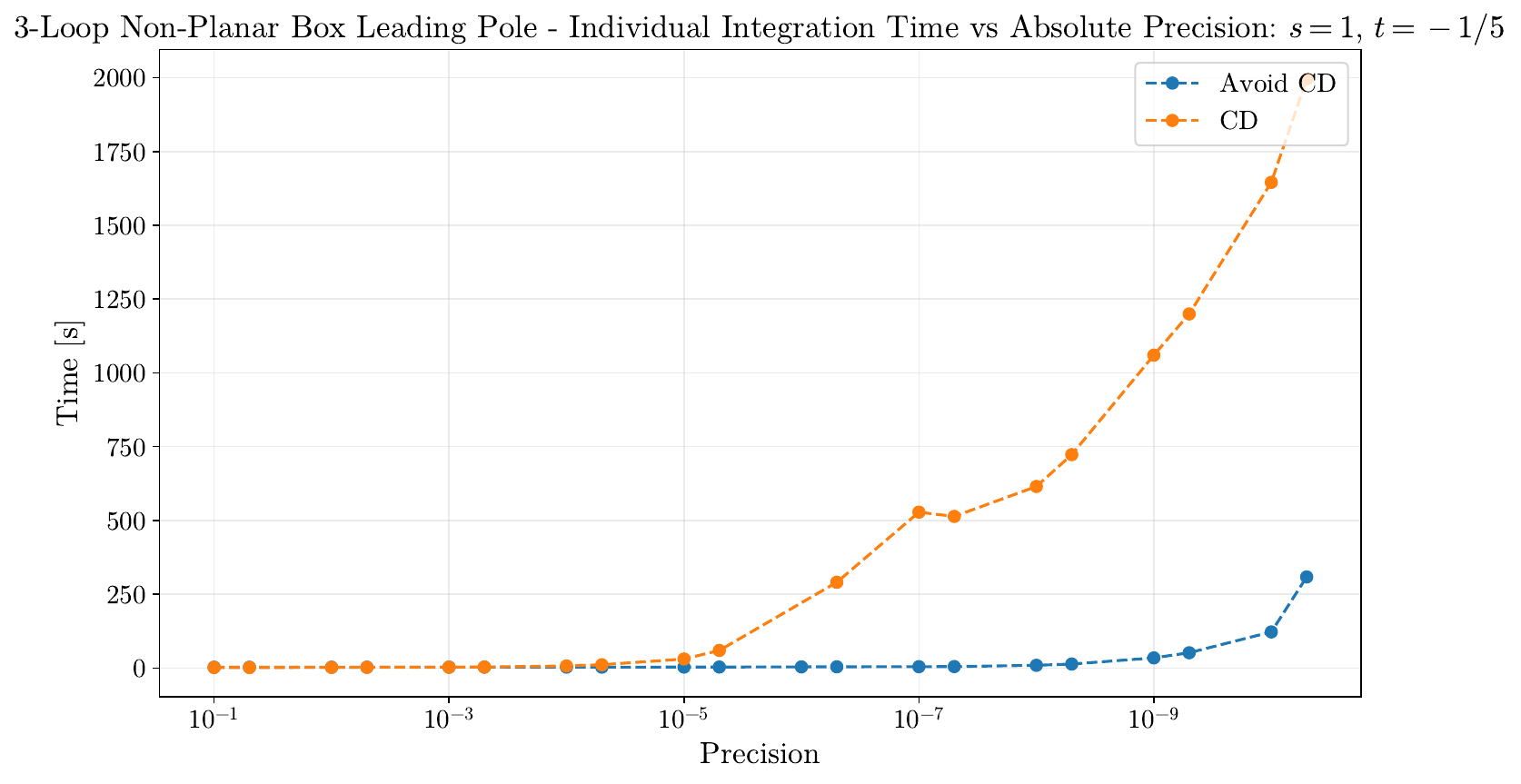}
    \caption{Timing comparison to obtain the leading pole ($1/\epsilon^4$) coefficient of the three-loop non-planar-box.}
    \label{fig:threeloopplot}
\end{figure}

The final example of massless integrals is the three-loop non-planar box with an $\mathscr{F}$ polynomial given by
\begin{equation}
    \mathscr{F}=-s \left(x_1 x_4-x_0 x_5\right)\left(x_3 x_6-x_2 x_7\right)-t\left(x_1 x_2 -x_0 x_3\right)\left(x_5 x_6-x_4x_7\right).
\end{equation}
As previously mentioned, this an example of a case where the integral has a leading Landau singularity for arbitrary kinematics \cite{sjpaper} and so the standard ``out-the-box'' implementation of contour deformation will inevitably fail. As explained in \cite{sjpaper}, this Landau singularity can be resolved by first linearising the singular hypersurface and then by considering all permutations of hierarchies for the Feynman parameters; finally, after taking into account symmetry, we are left with six independent integrals. For the regime $\mathcal{M}_{\mathrm{phys}}=\{(s,t)\in\mathbb{R}^2\ |\ s>|t|\land t<0\}$, we find that two of these six integrals would usually require contour deformation. Their $\mathscr{F}$ polynomials are
\begin{align}
    &\mathscr{F}_a=x_1x_3x_5x_7\left[-sx_0x_2+|t|\left(x_0+x_4\right)\left(x_2+x_4\right)\right]\\
    &\mathscr{F}_b=x_1x_3x_5x_7\left[sx_6\left(x_0+x_2+x_6\right)-|t|\left(x_0+x_6\right)\left(x_2+x_6\right)\right].
\end{align}
Through a more involved combination of splits and rescalings, we can express each of these integrals in terms of four others, resulting in a total of twelve integrals to compute (none of which require contour deformation):
\begin{equation}
    I=\sum\limits_{n_+=1}^{8}I^{+}_{n_+}+(-1-i\delta)^{-2-3\epsilon}\sum\limits_{n_-=1}^{4}I^{-}_{n_-}.
\end{equation}
Having verified the numerical result of this construction against the known analytic result \cite{Henn_2020,Bargie_a_2022}, we compared timings between numerically computing the six split integrals (post-resolution of the Landau singularity) using the usual contour deformation procedure and numerically computing the twelve integrals required to avoid contour deformation and these are shown in Fig.~\ref{fig:threeloopplot}.
We were not able to go beyond the leading pole ($1/\epsilon^4$) for the standard contour deformation setup whereas we could expand deeper in $\epsilon$ with our method for avoiding the procedure.

For the point $(s=1,t=-1/5)\in\mathcal{M}_{\mathrm{phys}}$, we state the precisions obtained after $\mathcal{O}\!\left(\text{mins}\right)$ with \pysecdec{} as well as the analytic result for comparison:
\begin{alignat}{2}    &I_\text{CD}&&=\ \left[8.340\textbf{55}-52.36\textbf{08}i\right]\epsilon^{-4}\ +\ \mathcal{O}(\epsilon^{-3})\\
&I_\text{Avoid CD}&&=\ \left[8.3400403920\textbf{28}\ \ -\ \ 52.35987755983\textbf{47}i\right]\epsilon^{-4}\ +\ \mathcal{O}(\epsilon^{-3})\\
&I_\text{analytic}&&=\ \left[8.34004039223768-52.35987755984493 i\right]\epsilon^{-4}\ +\ \mathcal{O}(\epsilon^{-3})
\end{alignat}
where the boldface digits denote the reported error on the final two stated digits preceding them.

\section{Massive Integrals}
Integrals with massive internal propagators (henceforth, ``massive integrals'') appear in a variety of phenomenologically-relevant amplitudes; for example, QCD corrections to processes involving massive quarks or electroweak corrections which contain a number of different mass scales resulting in integrals which often have no known analytic solution. In this scenario, numerical methods are essential -- the question therefore arises whether the method presented in this work extends to this class of integrals. The essential difference is that the $\mathscr{F}$ polynomial is modified by an extra term proportional to the $\mathscr{U}$ polynomial:
\begin{equation}
\mathscr{F}\left(\mathbf{x},\mathbf{s}\right)\ =\ \mathscr{F}_0\left(\mathbf{x},\mathbf{s}\right)\ +\ \mathscr{U}\left(\mathbf{x}\right)\sum\limits_{j=1}^{N}m_j^2x_j
\end{equation}
where $\mathscr{F}_0\left(\mathbf{x},\mathbf{s}\right)$ is the polynomial for the corresponding massless integral. This modification means that each individual Feynman parameter $x_j$ may appear quadratically in $\mathscr{F}$ as opposed to the massless case where, in each term, $x_j$ can appear linearly at most. The effect of this is that viable transformations can be difficult to deduce even for the most trivial of massive integrals. This motivates us to consider whether geometrical intuition can guide us in the right direction.

\begin{figure}[t]
    \centering
    \begin{subfigure}[t]{0.32\textwidth}
        \centering
        \begin{tikzpicture}[baseline=13ex,scale=1.0]
            %coordinates:
            \coordinate (x1) at (1, 2) ;
            \coordinate (x2) at (3, 2) ;
            %external momenta:
            \node (p1) at (0, 2) {$p$};
            \node (p2) at (4, 2) {};
            %jets:
            \draw[ultra thick,color=ForestGreen] (x1) -- (p1);
            \draw[ultra thick,color=ForestGreen] (x2) -- (p2);

            %edges:
            \draw[ultra thick,color=Purple] (x2) arc (0:180:1) node [midway,yshift=+10pt,color=Red] {\Large $m_1$};
            \draw[ultra thick,color=Purple] (x2) arc (0:-180:1) node [midway,yshift=-10pt,color=Red] {\Large $m_2$};

            %vertices:
            \draw[fill,thick,color=Blue] (x1) circle (1pt);
            \draw[fill,thick,color=Blue] (x2) circle (1pt);
        \end{tikzpicture}
    \end{subfigure}
    \hfill
    \begin{subfigure}[t]{0.32\textwidth}
        \centering
        \begin{tikzpicture}[baseline=14ex,scale=1.0]
            %coordinates:
            \coordinate (x1) at (1.1340, 1.4999) ;
            \coordinate (x2) at (2.8660, 1.5000) ;
            \coordinate (x3) at (2,3) ;
            %external momenta:
            \node (p1) at (0.2681, 0.9998) {$p_1$};
            \node (p2) at (3.7320, 1.0000) {$p_2$};
            \node (p3) at (2,4) {$p_3$};
            %jets:
            \draw[color=blue] (x1) -- (p1);
            \draw[ultra thick,color=ForestGreen] (x3) -- (p3);
            \draw[color=blue] (x2) -- (p2);
            %edges:
            \draw[ultra thick,color=Black] (x1) -- (x2);
            \draw[ultra thick,color=Black] (x2) -- (x3);
            \draw[ultra thick,color=Purple] (x3) -- (x1) node [midway,xshift=-10,color=Red] {\Large $m$};
            %vertices:
            \draw[fill,thick,color=Blue] (x1) circle (1pt);
            \draw[fill,thick,color=Blue] (x2) circle (1pt);
            \draw[fill,thick,color=Blue] (x3) circle (1pt);
        \end{tikzpicture}
    \end{subfigure}
    \hfill
    \begin{subfigure}[t]{0.32\textwidth}
        \centering
        \begin{tikzpicture}[baseline=13ex,scale=1.0]
            %coordinates:
            \coordinate (x1) at (1, 2) ;
            \coordinate (x2) at (3, 2) ;
            %external momenta:
            \node (p1) at (0, 2) {$p$};
            \node (p2) at (4, 2) {};
            %jets:
            \draw[ultra thick,color=ForestGreen] (x1) -- (p1);
            \draw[ultra thick,color=ForestGreen] (x2) -- (p2);

            %edges:
            \draw[ultra thick,color=Purple] (x2) arc (0:180:1) node [midway,yshift=+10pt,color=Red] {\Large $m_1$};
            \draw[ultra thick,color=Purple] (x2) arc (0:-180:1) node [midway,yshift=+10pt,color=Red] {\Large $m_3$};
            \draw[ultra thick,color=Purple] (x1) -- (x2) node [midway,yshift=+10pt,color=Red] {\Large $m_2$};

            %vertices:
            \draw[fill,thick,color=Blue] (x1) circle (1pt);
            \draw[fill,thick,color=Blue] (x2) circle (1pt);
        \end{tikzpicture}
    \end{subfigure}
    \caption{The massive integrals to which the method has been successfully applied (\textbf{left} \& \textbf{centre}) and the massive sunrise integral (\textbf{right}) which we are currently investigating. The colour scheme matches Fig.~\ref{fig:masslessdiags} and, additionally, purple lines denote massive internal propagators with their masses given in red.}
    \label{fig:massivediags}
\end{figure}
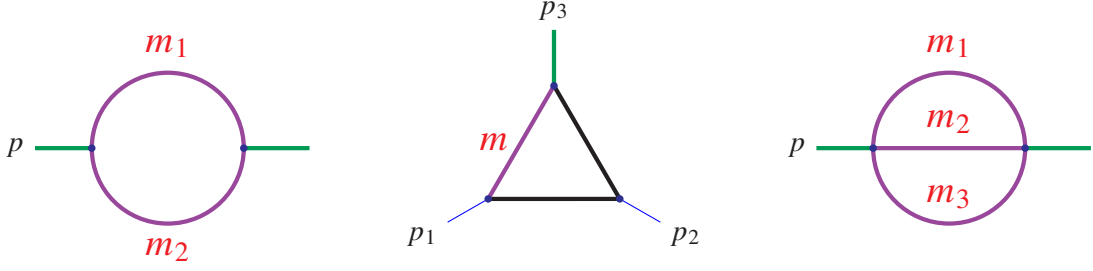

We have successfully employed the method on the massive bubble (which we describe in detail in Section~\ref{sec:massbub}) and triangle (both shown in Fig.~\ref{fig:massivediags}) and current work involves the investigation of the sunrise integral (also in Fig.~\ref{fig:massivediags}) which is known to be the simplest Feynman integral which cannot be expressed in terms of multiple polylogarithms but rather elliptic integrals.

% \begin{figure}[t]
%     \centering
%     \includegraphics[width=1.0\textwidth]{massivediags.png}
%     \caption{The massive integrals to which the method has been successfully applied (\textbf{left} \& \textbf{centre}) and the massive sunrise integral (\textbf{right}) which we are currently investigating. The colour scheme matches Fig.~\ref{fig:masslessdiags} and, additionally, purple lines denote massive internal propagators with their masses given in red.}
%     \label{fig:massivediags}
% \end{figure}

\subsection{Massive Bubble}\label{sec:massbub}
\begin{figure}[t]
    \centering
    \begin{tikzpicture}
        \node[anchor=south west,inner sep=0] (img) at (0,0) {\includegraphics[width=0.5\textwidth]{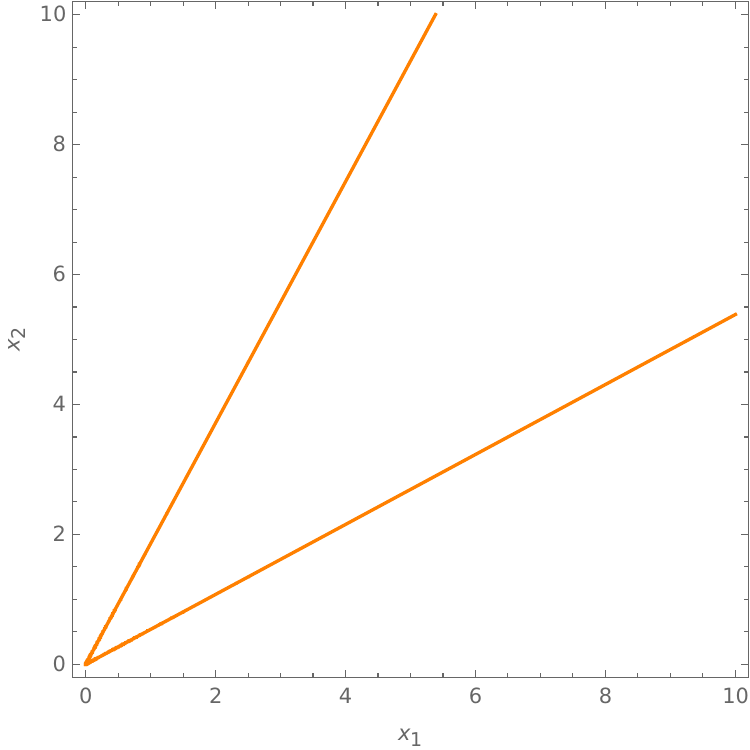}};
        \begin{scope}[x={(img.south east)}, y={(img.north west)}]
            % % Draw grid lines
            % \draw[help lines, line width=0.1pt, blue] (0,0) grid (1,1);
            % % Add ticks to the x-axis
            % \foreach \x in {0, 0.1, ..., 1} {
            %     \draw (\x, 0) -- (\x, -0.02) node[anchor=north] {\tiny \x};
            % }
            % % Add ticks to the y-axis
            % \foreach \y in {0, 0.1, ..., 1} {
            %     \draw (0, \y) -- (-0.02, \y) node[anchor=east] {\tiny \y};
            % }
            % Add the equation text
            \node at (0.7,0.7) {$x_2 = \frac{1 \pm \beta}{1 \mp \beta} x_1$};
            % Add the arrow
            \draw [->, thick] (0.76,0.64) -- (0.83,0.52);
            \draw [->, thick] (0.66,0.74) -- (0.52,0.83);
        \end{scope}
    \end{tikzpicture}
    \caption{Plot of the variety (in orange) of the $\widetilde{\mathscr{F}}$ polynomial for the massive bubble which separates three regions in the positive quadrant of $\mathbb{R}^{2}$.}
    \label{fig:bubblevariety}
\end{figure}
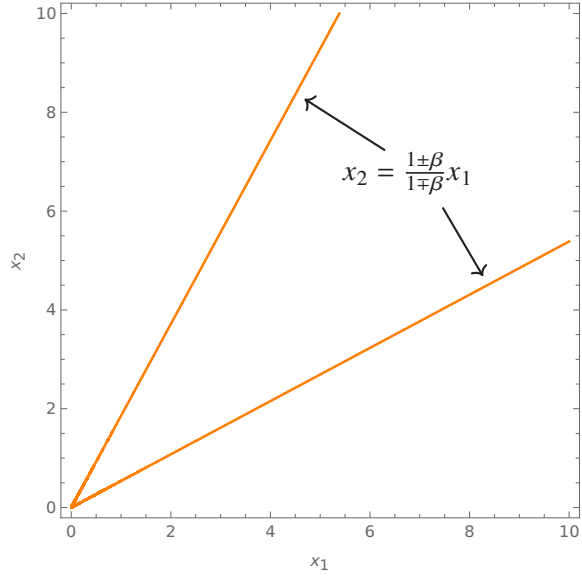

The $\mathscr{F}$ polynomial of the massive bubble is given by
\begin{equation}
    \mathscr{F}=-p^2x_1x_2+\left(x_1+x_2\right)\left(m_1^2x_1+m_2^2x_2\right).
\end{equation}
We define $\beta^2:=\frac{p^2-\left(m_1+m_2\right)^2}{p^2-\left(m_1-m_2\right)^2}\in\left[0,1\right)$ and scale out the dimension of $\mathscr{F}$ with the transformations $x_i\rightarrow\frac{x_i}{m_i}$ to instead analyse the dimensionless polynomial
\begin{equation}
    \widetilde{\mathscr{F}}=x_1^2+x_2^2-2\frac{1+\beta^2}{1-\beta^2}x_1 x_2.
\end{equation}
It is illuminating to plot the variety of $\widetilde{\mathscr{F}}$ and this is given in Fig.~\ref{fig:bubblevariety}.
We see that there are three distinct regions separated by the codimension-1 zero hypersurface of $\widetilde{\mathscr{F}}$, two of which are defined by $\widetilde{\mathscr{F}}$ being positive, the remaining region defined by $\widetilde{\mathscr{F}}$ being negative. 

We can construct transformations which directly send the variety to the integration boundary. For example, for the upper-left region of the plot, we demand of the transformed variables \{$y_i$\} that the $y_2$-axis coincides with the $x_2$-axis and the $y_1$-axis coincides with the solution line $x_2=\frac{1+\beta}{1-\beta}x_1$. Along with the constraint that points within this region get mapped to the positive quadrant in the new $y_i$ variables, this uniquely defines the transformation:
\begin{equation}
    y_1\overset{!}{=}x_1,y_2\overset{!}{=}x_2-\frac{1+\beta}{1-\beta}x_1\Rightarrow x_1\rightarrow y_1,x_2\rightarrow y_2+\frac{1+\beta}{1-\beta}y_1.
\end{equation}

We have verified that the resulting construction $I=I_1^++I_2^++(-1-i\delta)^{-\epsilon}I^-_1$ replicates the analytic result by integrating it both numerically with \pysecdec{} \textit{and} analytically (as we can easily calculate the constituent integrals in the sum symbolically, for example, using \textsc{Mathematica}).

\section{Conclusions \& Outlook}
As we have shown, this method allows us to avoid contour deformation (and its associated computational inefficiencies) in regions of kinematic space which would ordinarily require it. We demonstrate that the time to evaluate multi-loop integrals numerically can be improved by a factor of one thousand or more in selected cases and the ability to expand deeper in the dimensional regulator $\epsilon$ has also been observed for a particularly pathological three-loop example. We have also conveyed that geometry can play a role in deriving the necessary transformations of the Feynman parameters to enable the avoidance of contour deformation, for example, in massive integrals where they are considerably more difficult to identify.

Current and future work is directed towards utilising this intuition attained by visualising the problem geometrically to move away from arbitrary shift and rescaling transformations (derived from a trial-and-error approach) and towards a general algorithm to define canonical transformations. There is potential insight to be gained from techniques in the fields of algebraic and tropical geometry by directly considering the variety of the Symanzik polynomial $\mathscr{F}$ and using computational tools for the resolution of singularities \cite{Hironaka}. 

Understanding how to apply this method to two- and even three-loop massive integrals could lead to very significant time improvements, perhaps even better than the orders of magnitude improvements we have already presented in these proceedings for massless integrals. Moreover, applying the method to further multi-loop massless integrals may allow for (currently-impossible) numerical cross-checks of analytic or semi-analytic results.

The long-term objective of this work is to implement this technique in numerical loop integration packages like \pysecdec{} to be used in the efficient calculation of QFT amplitudes where, along with the integration-by-parts reduction to master integrals, multi-loop integral evaluation remains a major bottleneck.

\acknowledgments

We thank the members of the \pysecdec{} collaboration, Gudrun Heinrich, Matthias Kerner, Vitaly Magerya, and Johannes Schlenk, for discussions and insightful input.
This research was supported in part by the Deutsche Forschungsgemeinschaft (DFG, German Research Foundation) under grant 396021762 - TRR 257, and by the UK Science and Technology Facilities Council under contract ST/T001011/1. 
SJ is supported by a Royal Society University Research Fellowship (Grant URF/R1/201268).

\bibliographystyle{JHEP}
\bibliography{main}

\end{document}